# Narrow Linewidth Distributed Feedback Lasers Utilizing Distributed Phase Shift

Yiming Sun, Bocheng Yuan, Xiao Sun, Simeng Zhu, Yizhe Fan, Mohanad Al-Rubaiee, John H. Marsh, Stephen J. Sweeney and Lianping Hou

*James Watt School of Engineering, University of Glasgow, Glasgow, G12 8QQ, United Kingdom*

**Abstract: This study proposes and experimentally demonstrates a distributed feedback (DFB) laser with a distributed phase shift (DPS) region at the center of the DFB cavity. By modeling the field intensity distribution in the cavity and the output spectrum, the DPS region length and phase shift values have been optimized. Experimental comparisons with lasers using traditional π-phase shifts confirm that DFB lasers with optimized DPS lengths and larger phase shifts (up to 15π) achieve stable single longitudinal mode operation over a broader current range, with lower threshold current, higher power slope efficiency, and a higher side mode suppression ratio (SMSR). Furthermore, the minimum optical linewidth is reduced significantly, from 1.3 MHz to 220 kHz.**

Narrow linewidth lasers have widespread applications in coherent optical communications, spectroscopy, metrology, and various other fields [1]. Among the available laser technologies for narrow linewidth lasers, distributed feedback (DFB) lasers have become prominent due to their unique distributed feedback mechanism. This mechanism facilitates narrow linewidth, single longitudinal mode emission, wavelength stability, and strong integration capabilities. To ensure single longitudinal mode (SLM) operation, conventional practice typically involves incorporating a quarter-wavelength phase shift section into the middle of the DFB laser cavity [2]. However, the quarter-wavelength phase-shift section creates a point defect within the cavity, leading to an uneven intracavity field distribution. At high optical power levels, this uneven distribution exacerbates spatial hole burning (SHB), leading to reduced overall efficiency, degraded mode stability, and an increase in optical linewidth [3]. To mitigate the adverse effects of this point defect, a technique known as multi-phase-shift (MPS) gratings has been employed in which the phase shift is distributed across different locations within the cavity [4]. However, for optimal performance, the number and positioning of the phase shifts must be precisely calculated and optimized. This increases the complexity of the design process. Another solution is to use corrugation-pitch-modulated (CPM) gratings, which feature a phase-adjusting region where the π phase shift is evenly distributed across the phase-adjusting region. However, modifying the grating period not only alters the Bragg conditions but also introduces manufacturing challenges, as achieving the required precision for producing grating periods with extremely small differences (<0.1 nm) is difficult [5]. To overcome these challenges, more complex structures have been proposed, such as equivalent CPM gratings [6, 7] and asymmetric three corrugation-pitch-modulated DFB gratings [8].

In this work, we propose and demonstrate a novel grating structure, termed the Distributed Phase Shift (DPS) grating, which maintains a constant grating period across both DPS and non-DPS regions. Instead of varying the grating period, we introduce a linear phase adjustment within each grating period, ensuring that the total phase shift in the DPS region equals π, 3π, 5π, 7π, and 15π, respectively. This design effectively mitigates the spatial hole burning (SHB) effect, typically caused by the quarter-wavelength phase shift in conventional DFB lasers and achieves narrow linewidth output.

Our approach extends the quarter-wavelength phase shift (π) from π to as much as 15π, distributing it across the DPS region, which was initially located at the cavity center. This method im-proves the uniformity of the optical field distribution and manufacturing resolution. Our solution is simpler to design and implement in grating structures, while offering enhanced performance in terms of optical linewidth and side-mode suppression ratio (SMSR) compared to conventional quarter-wavelength phase shift DFB lasers. Furthermore, it maintains stable optical linewidth performance over a wider operating current range compared to previously reported techniques, such as MPS and CPM [4, 5].

For a detailed comparison, Fig. 1(a) presents the schematic structures of the various grating types discussed above. In contrast to a uniform grating (UG) structure, the quarter-wavelength phase-shifted grating incorporates a π phase shift at the center of the cavity, corresponding to a half-cycle phase shift. Our novel DPS grating combines the advantages of both CPM and MPS gratings. The DPS grating is similar to the CPM grating, featuring a specific region ($L_P$) over which the π phase shift is distributed. However, its overall configuration closely resembles that of a uniform grating. This similarity is due to the unique grating design, which maintains the original grating period while gradually introducing phase shifts along the designated DPS region, with each phase shift occurring only at the beginning of each original grating period. Additionally, we extend the total phase shift in the DPS region from the conventional π to 15π and analyze the impact of the DPS length and phase shift value on the DFB laser's intensity uniformity within the cavity, as well as its SMSR.

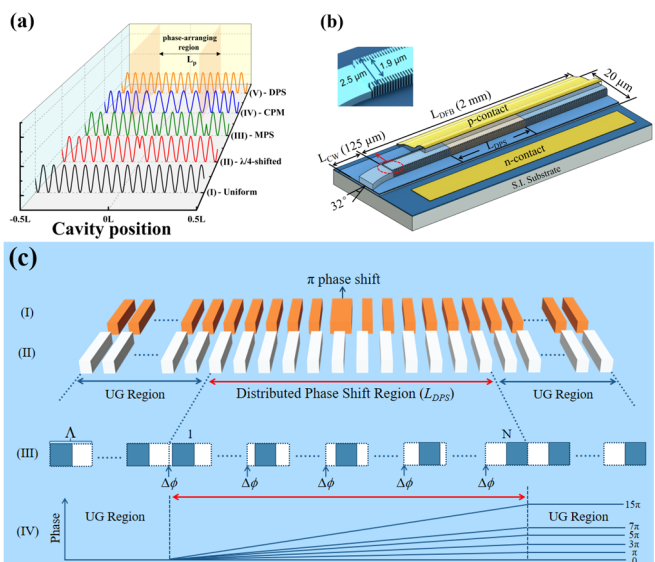

Fig. 1. (a) Schematic structure for various types of grating: (I) uniform grating, (II) quarter-wavelength phase-shift grating, (III) multi-phase-shift (MPS) grating, (IV) corrugation-pitch-modulated (CPM) gratings, (V) Distributed Phase Shift (DPS) grating. (b) Schematic of the DFB structure based on the DPS grating, with the inset showing the dimensions of the ridge waveguide featuring sidewall gratings. (c) Grating schematic of (I) quarter-wavelength phase-shift grating, (II) DPS grating, (III) the π phase change within the DPS grating, and (IV) designed total phase shift of π, 3π, 5π, 7π, and 15π in the DPS region.

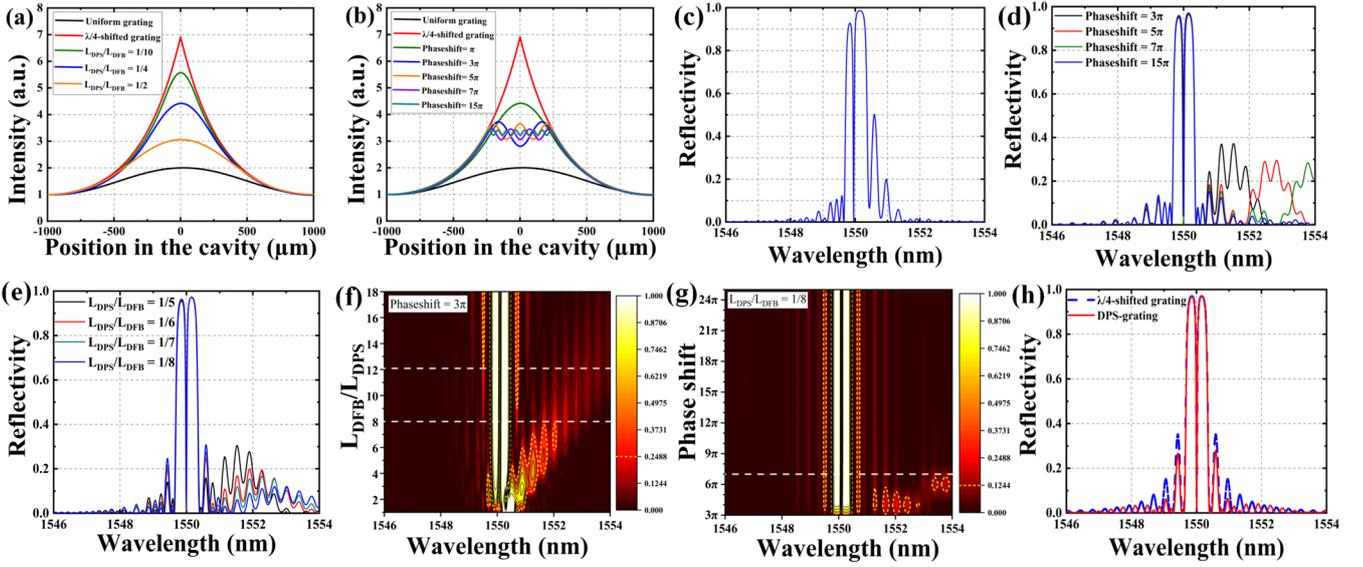

Fig. 2. Calculated light field distribution along the cavity for different types of DPS gratings compared to uniform grating and traditional π phase shift grating: (a) different values of $L_{DPS}/L_{DFB}$ with a fixed π phase shift, (b) different phase shifts with a fixed $L_{DPS}/L_{DFB} = 1/4$; Calculated reflection spectra of (c) DPS grating with a π phase shift in the DPS region and $L_{DPS}/L_{DFB} = 1/4$, (d) DPS grating with 3π, 5π, 7π and 15π phase shifts in the DPS region and $L_{DPS}/L_{DFB} = 1/4$, (e) 3π DPS grating with $L_{DPS}/L_{DFB} = 1/5, 1/6, 1/7$ and $1/8$; Calculated 2D optical spectra (f) vs $L_{DFB}/L_{DPS}$ (1-18) with the phase shift fixed at 3π, (g) vs the phase shifts (3π-25π) with $L_{DPS}/L_{DFB} = 1/8$; (h) Calculated reflection spectra of 15π DPS grating with $L_{DPS}/L_{DFB} = 1/8$ compared with traditional π phase shift grating.

Fig. 1(b) illustrates a sidewall grating DFB laser featuring a DPS grating structure. This DFB laser utilizes a three-quantum-well (QW) AlGaInAs active layer, complemented by a far-field reduction layer (FRL) situated beneath the active layer on a semi-insulating (S.I.) InP substrate. The detailed epilayer structure is provided in [9]. The top ridge waveguide is 2.5 μm wide and 1.92 μm high, with a grating recess depth of 0.3 μm. The second ridge, incorporating the active layer with three QWs, is 20 μm wide and 608 μm high, and is designated for n-contact deposition. To reduce the optical overlap with the QWs, reduce the vertical divergence angle and improve the coupling efficiency to single mode fiber (SMF), a 160 nm FRL is placed beneath the active layer and 400 nm InP spacer layer. The total cavity length of the DFB laser ($L_{DFB}$) is 2 mm, with the DPS region centered in the grating structure, spanning a length of $L_{DPS}$. To minimize end-facet reflections and reduce Fabry–Perot cavity effects, a 125 μm long unpumped curved waveguide with a radius of 233.3 μm and a 32° angle is incorporated at one of the laser's output facets.

The modulation scenario of the DPS grating is shown in Fig. 1(c). The DPS grating maintains the same uniform grating period $Λ$ across both DPS and non-DPS regions. The DPS region starts at the end of the previous grating period in the UG region, introducing the initial phase shift $Δϕ$, where $Δϕ = π/N$, with $N$ representing the number of grating periods in the DPS region, calculated by $N = L_{DPS}/Λ$; it also signifies the total number of phase shifts applied in the DPS region, then, each phase shift is introduced at the commencement of each original grating period. Therefore, the total phase shift produced in the DPS region, π, replaces the single π phase shift found in traditional π phase shift gratings. The incremental phase shifts do not change the dimensions of the grating period; they only determine the position of the grating within each period. Furthermore, the total phase shift can be set to 3π, 5π, 7π, or 15π, with the phase shift increment defined as $Δϕ = nπ/N$, where n = 3, 5, 7, or 15, as illustrated in part (IV) of Fig. 1(c).

The relationship between the linewidth ($Δν$) and the cavity parameters is described by the Schawlow-Townes linewidth formula, as shown in Equation (1), which accounts for the quantum mechanical noise due to spontaneous emission within the laser cavity [10]:

$$Δν = \frac{ν_g^2 hνgn_{sp}α_{th}(1+α^2)}{4πP_0} \qquad (1)$$

where $P_0$ is the output power, $hν$ is the photon energy, $ν_g$ is the group velocity, $g$ is the gain, $n_{sp}$ refers to the spontaneous emission factor, $α$ is the detuning parameter and $α_{th}$ is the net amplitude threshold gain. In the DFB laser, $α_{th}$ can be approximated as $π^2/κ^2 L_{DFB}^3$ [11]. Increasing $κL_{DFB}$ reduces the linewidth but can cause mode instability due to SHB. The linewidth is inversely proportional to the cavity length ($L_{DFB}$). As $L_{DFB}$ increases, the threshold carrier density decreases, leading to reductions in both $α$ and $n_{sp}$. This suggests that further linewidth reduction is possible. Therefore, the most effective strategy to minimize linewidth is to increase the cavity length ($L_{DFB}$) while maintaining $κL_{DFB}$ at its optimal value [12]. As mentioned above, we use a 3-QW active layer with an FRL structure, a grating recess depth of 0.3 μm, and an $L_{DFB}$ of 2 mm. This configuration yields an optical confinement factor of 2.19%, a $κL_{DFB}$ value of 1.25 for conventional quarter-wavelength phase-shift gratings, and a $κL_{DFB}$ value of approximately 1.0 for DFB lasers with a DPS structure.

The distribution of the light field within the cavity is influenced by the design of the grating structure. Using the transfer matrix method (TMM), the light field distributions for various grating structures are illustrated in Fig. 2. In the calculations, the effective index is set to 3.19 and the grating period to 243 nm, ensuring an output wavelength of 1550 nm. The cavity length $L_{DFB}$ is set to 2 mm.

Fig. 2(a) shows the calculated light intensity distribution of a uniform grating, traditional π phase-shifted grating, and DPS grating with different $L_{DPS}/L_{DFB}$. Compared to a uniform grating, the traditional π phase-shifted grating exhibits a pronounced intensity peak at the center, significantly disrupting the flatness of the optical field distribution. With the introduction of a DPS grating with π phase shift, while $L_{DPS}/L_{DFB} = 1/10$, it can be seen from the figure that the maximum value in the cavity center is attenuated. As $L_{DPS}/L_{DFB}$ is increased from 1/10 to 1/4, the light field of the cavity is significantly smoothed, approaching the flatness level of the uniform grating.

However, two issues remain. The first problem is that, based on the above assumptions, when $L_{DPS}/L_{DFB} = 1/4$, each phase shift corresponds to a difference of only 0.25 nm in grating movement, which exceeds the typical resolution limit of 0.5 nm in electron beam lithography (EBL). The second issue is the instability in the output spectrum caused by a single π phase shift in the DPS grating.

Fig. 2(c) shows the reflection spectra of a DPS grating with a single π phase shift at the DPS region and $L_{DPS}/L_{DFB} = 1/4$. The laser output is not centered at the main wavelength, and a significant side mode effect is also present. To solve this problem, the multi-π phase shift has been introduced. Fig. 2(d) shows the reflection spectra of DPS gratings with 3π, 5π, 7π and 15π phase shifts in the DPS region. As the value of phase shift increases, the extra side modes are suppressed and move away from the output central wavelength. Based on this, we simulated the light field distribution and found that as the phase shift value increases, the light field distribution in the DPS region becomes flatter, eventually approaching a smooth straight line, as shown in Fig. 2(b). Fig. 2(e) illustrates the impact of DPS length on the reflection spectrum with a fixed 3π phase shift. As the ratio of the DPS region to the total cavity length decreases from 1/5 to 1/8, additional side modes are more effectively suppressed and move further away from the main lasing longitudinal mode.

To optimize the performance of DPS DFB lasers, the appropriate DPS region length and phase shift size need to be adjusted. Fig. 2(f) shows the simulated two-dimensional (2D) optical spectrum versus $L_{DFB}/L_{DPS}$ (1-18) with phase shift fixed at 3π. It is found that, while $L_{DFB}/L_{DPS}$ is larger than 8, the reflection intensities of other modes in the spectrum are much smaller than that of the first side lobe. Fig. 2(g) shows the simulated two-dimensional (2D) optical spectrum versus the value of phase shift (ranging from 3π to 25π) with $L_{DPS}/L_{DFB} = 1/8$. As the phase shift increases, the reflectivity of other modes decreases further and, when the phase shift exceeds 7π, the reflectance spectrum closely resembles that of a traditional π-phase-shifted grating. Fig. 2(h) shows the traditional phase-shifted grating and DPS grating for 15π and $L_{DPS}/L_{DFB} = 1/8$. Under this condition, we find that the side modes of the DPS grating with 15π phase shift are effectively suppressed. This suppression is achieved through the guided mode resonance between the DPS region and the two adjacent uniform gratings. The DPS ensures that the primary mode is favored over side modes, leading to stable SLM operation and a high SMSR. More importantly, compared to the initial design with $L_{DPS}/L_{DFB} = 1/4$ and a π phase shift, the required manufacturing resolution has been significantly improved from 0.25 nm to 7.5 nm.

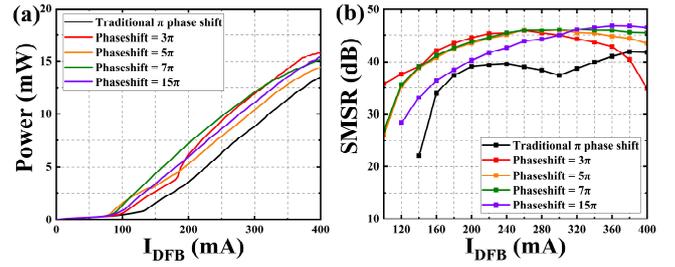

Fig.3. (a) Typical $P$-$I_{DFB}$ curves of traditional π phase shift grating and 3π, 5π, 7π and 15π phase shifted DPS gratings with $L_b/L = 1/8$, (b) SMSRs versus $I_{DFB}$ for traditional π phase shift grating and DPS gratings with 3π, 5π, 7π, and 15π phase shifts, with $L_b/L = 1/8$.

The device fabrication process followed a procedure similar to that described in [13], with the exception that the p- and n-contacts were defined concurrently on the same side of the wafer. The fabricated devices included traditional π phase-shifted gratings and DPS devices with varying lengths and phase shifts in the DPS region, all co-located on the same wafer and fabricated simultaneously. After fabrication, the laser chip was cleaved into bars with both facets left uncoated. Each laser bar was mounted epilayer-up onto a copper heat sink, which was placed on a Peltier cooler. The heat sink temperature was set to 20 °C, and the devices were tested under continuous-wave (CW) conditions.

The power-current (P-I) characteristics of different types of grating devices are shown in Fig. 3(a). The threshold current of the traditional π phase-shifted laser is 122 mA and the threshold current of DPS grating devices is around 85 mA. The results showed a significant reduction in threshold current for the DPS grating devices, attributed to the uniform light field distribution within the cavity, which enhances the overlap between the optical field and the gain medium [14]. For traditional phase-shifted lasers, the output power reaches its maximum of 13 mW at a DFB current ($I_{DFB}$) of 400 mA. In contrast, all DPS grating devices demonstrate higher output power at the same current. Without considering mode hopping, as the phase shift value increases, the output power also rises, with the highest output of 15.5 mW achieved by the 15π DPS grating device at $I_{DFB}$ =400 mA. This confirms that a larger phase shift in the DPS improves the uniformity of the optical field distribution, reduces phase noise, and enhances the feedback mechanism within the laser, resulting in a lower threshold current and higher slope efficiency compared to conventional DFB lasers with a π-phase shift.

The SMSRs of traditional π phase-shifted and DPS grating devices with 3π, 5π, 7π and 15π versus $I_{DFB}$ are shown in Fig. 3(b). For the traditional π phase-shifted device, the SMSRs fluctuate around $I_{DFB}$=290 mA due to mode-hopping, with the maximum SMSR value reaching 41 dB at $I_{DFB}$=380 mA. By contrast, multi-π phase shift DPS grating devices exhibit more stable and higher SMSRs. The highest SMSR value achieved is 47 dB for the 15π phase shift DPS grating device, which operates in a stable SLM across $I_{DFB}$ range from 200 mA to 400 mA with an SMSR > 40 dB.

Fig. 4 shows a two-dimensional (2D) optical spectrum versus $I_{DFB}$ (0-450 mA) for different types of lasers. The spectrum of the traditional π phase-shifted grating is shown in Fig. 4(a), as the injection current is increased, there is a stable SLM output, and a mode hop appears at $I_{DFB}$=280mA. The measured $\kappa L_{DFB}$ value is 1.25. Fig. 4(b) shows the optical spectrum of a DPS grating with 3π and $L_{DPS}/L_{DFB} = 1/5$; it shows that when $I_{DFB}$>280mA, the output exhibits serious mode hopping and multi-mode phenomena, which is consistent with the simulation results in Fig. 2(e). The measured $\kappa L_{DFB}$ value is 0.98. Fig. 4(c) shows the optical spectrum of a DPS grating with 3π and $L_{DPS}/L_{DFB}$ = 1/8. Under these conditions, the SLM characteristics of the output light are significantly improved. The measured $\kappa L_{DFB}$ value is 1.1. Fig. 4(d), (e), and (f) show the spectra for DPS gratings with $L_{DPS}/L_{DFB}$ = 1/8 and phase shifts of 5π, 7π and 15π, respectively. Very stable SLM operation is observed from the threshold current (85 mA) up to 450 mA, without any mode-hopping. The average current-induced wavelength redshift coefficient is around 0.0075 nm/mA. The measured $\kappa L_{DFB}$ value is also 1.1, as all structures have the same DPS length, with $L_{DPS}/L_{DFB}$=1/8.

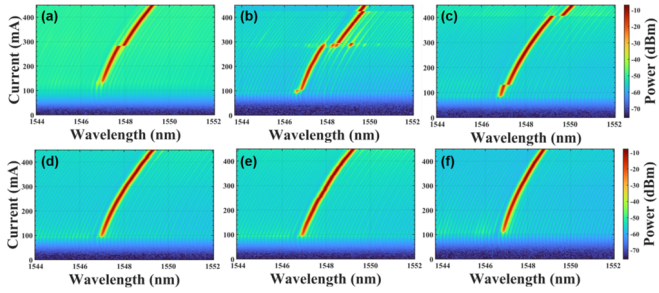

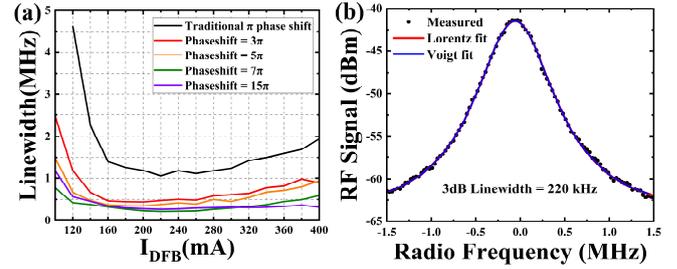

Fig. 4. 2D optical spectra vs $I_{DFB}$ for (a) traditional π phase shift grating, (b) 3π DPS grating with $L_{DPS}/L_{DFB}$ = 1/5, (c) 3π DPS grating with $L_{DPS}/L_{DFB}$ = 1/8, (d) 5π DPS grating with $L_{DPS}/L_{DFB}$ = 1/8, (e) 7π DPS grating with $L_{DPS}/L_{DFB}$ = 1/8, (f) 15π DPS grating with $L_{DPS}/L_{DFB}$ = 1/8.

Fig. 5. (a) Measured optical linewidth versus $I_{DFB}$ for traditional π phase shift grating and DPS gratings with 3π, 5π, 7π, and 15π phase shifts, with $L_{DPS}/L_{DFB}$ = 1/8, (b) RF beat note signal fitted to Lorentzian (red dot) and Voigt (blue dot) profiles for the narrowest achieved linewidth, with a FWHM of 220 kHz, obtained using the 7π DPS grating device at $I_{DFB}$ = 260 mA.

The linewidth of the fabricated devices was measured using the delayed self-heterodyne technique, employing a 12.5 km long single-mode fiber and an 80-MHz acoustic-optic modulator [15]. During the measurement, the resolution of the electrical signal analyzer (ESA) was kept at 20 kHz, and the measured radio frequency (RF) spectra were fitted to Lorentz and Voigt profiles to calculate the −3-dB linewidth of the DFB laser, which is the typical line shape expected from DFB semiconductor lasers. The measured linewidth as a function of $I_{DFB}$ is shown in Fig. 5(a). The traditional π phase-shifted devices exhibit typical linewidth-current characteristics: as $I_{DFB}$ increases from 120 mA to 160 mA, the spectral linewidth narrows from 4.6 MHz to 1.45 MHz, before broadening again at approximately 280 mA. In contrast, the DPS grating devices achieve narrow linewidth characteristics. As the equivalent phase shift in the DPS region increases, linewidth broadening is effectively suppressed. When the phase shift is set to 15π, no evident linewidth broadening is observed within the $I_{DFB}$ range of 120 mA to 400 mA. The experimental results indicate that the minimum linewidth does not continue to decrease when the equivalent phase shift reaches 15π. This is because, as the phase shift further increases, the light field distribution in the DPS region tends to flatten into a straight line. Based on the results, we find that the DPS DFB laser maintains stable optical linewidth performance over a wider operating current range compared to previously reported techniques, such as MPS and CPM [4, 5]. Fig. 5(b) shows the RF beat note signal fitted to the Lorentz (red dot) and Voigt (blue dot) profiles for the narrowest achieved linewidth signal (220 kHz) with the 7π DPS grating device at $I_{DFB}$ = 260 mA.

In conclusion, compared to conventional π-phase-shift DFB lasers, DFB lasers utilizing the DPS technique achieve a more uniform optical field distribution within the laser cavity. This results in narrower linewidth output, higher output power, enhanced SMSR and improved SLM stability. The DPS approach is essential for reducing phase noise, increasing the differential quantum efficiency and ensuring mode stability by stabilizing the feedback mechanism. This technology paves the way for the development of high-power, narrow-linewidth, and stable DFB lasers that meet the rigorous demands of applications in coherent optical communication, spectroscopy, and metrology.

This work was supported by the U.K. Engineering and Physical Sciences Research Council under grant EP/R042578/1. We would like to acknowledge the staff of the James Watt Nanofabrication Centre at the University of Glasgow for their help in fabricating the devices.

## AUTHOR DECLARATIONS
### Conflict of Interest

The authors declare no conflict of interest.

### Author Contributions

Yiming Sun: Conceptualization (equal); Methodology (lead); Writing – original draft (lead). Bocheng Yuan: Conceptualization (equal); Writing – review & editing (equal). Xiao Sun: Conceptualization (equal); Writing – review & editing (equal). Simeng Zhu: Writing – review & editing (equal). Yizhe Fan: Writing – review & editing (equal). Mohanad Al-Rubaiee: Writing – review & editing (equal). John H. Marsh: Funding acquisition (equal), Writing – review & editing (equal). Stephen J. Sweeney: Writing – review & editing (equal). Lianping Hou: Supervision (lead), Funding acquisition (lead); Writing – review& editing (equal).

## DATA AVAILABILITY

The data that support the findings of this study are available from the corresponding author upon reasonable request.

## List of References with titles